\documentclass[aps,prl,twocolumn,superscriptaddress]{revtex4}

\usepackage{graphicx}
\graphicspath{\Users\tanjacuk\Documents\PressureWork\RamanPRL}
\pdfoutput=1
\begin{document}


\title{Uncovering a pressure-tuned electronic transition in  Bi$_{1.98.}$Sr$_{2.06}$Y$_{0.68}$Cu$_{2}$O$_{8+\delta}$
using Raman scattering and x-ray diffraction}


\author{T. Cuk}
\affiliation{Departments of Physics, Applied Physics and Stanford
Synchrotron Radiation Laboratory, Stanford University, Stanford,
CA 94305}
\author{V. Struzhkin}
\affiliation{Geophysical Laboratory, Carnegie Institution of
Washington, Washington, DC 20015}
\author{T.P. Devereaux}
\affiliation{Dept. of Physics University of Waterloo Waterloo, ON
CANADA, N2L 3G1}
\author{A. Goncharov}
\affiliation{Geophysical Laboratory, Carnegie Institution of
Washington, Washington, DC 20015}
\author{C. Kendziora}
\affiliation{Naval Research Laboratory, Code 6365, Washington, DC
20375}
\author{H. Eisaki}
\affiliation{Nanoelectronics Research Institute, National
Institute of Advanced Industrial Science and Technology, 1-1-1
Central 2, Umezono, Tsukuba, Ibaraki, 305-8568, Japan}
\author{H. Mao}
\affiliation{Geophysical Laboratory, Carnegie Institution of
Washington, Washington, DC 20015}
\author{Z.X. Shen}
\affiliation{Departments of Physics, Applied Physics and Stanford
Synchrotron Radiation Laboratory, Stanford University, Stanford,
CA 94305}



\begin{abstract}
We report pressure tuned Raman and x-ray diffraction data of
Bi$_{1.98}$Sr$_{2.06}$Y$_{0.68}$Cu$_{2}$O$_{8 + \delta}$ revealing a
critical pressure at 21 GPa with anomalies in six physical
quantities:  electronic Raman background, electron-phonon coupling
$\lambda$ , spectral weight transfer from high to low frequency,
density dependent behaviour of phonon and magnon frequencies, and
a compressibility change in the c-axis.  For the first time in a cuprate, mobile charge carriers,
lattice, and magnetism all show anomalies at a distinct critical pressure in the same
experimental setting. Furthermore, the Raman spectral changes are similar to 
that seen traversing the superconducting dome with doping, suggesting that the 
critical pressure at 21 GPa is related to the much discussed critical point at optimal doping.  \end{abstract}


\maketitle

One of the most outstanding questions in
condensed matter physics concerns the emergence of high temperature
superconductivity via doping a Mott insulator. The phase diagram of
the cuprates as a function of doping offers many clues.
The general phenomenology of the high-Tc phase diagram divides it
into distinct under-doped and over-doped regions characterized by,
respectively, insulating behaviour\cite{1,2}, gapped Fermi
surfaces\cite{3, 4, 5} charge, spin and current
order\cite{6,7,8,9,10,11} and fluctuating superconductivity
\cite{12,13} on the one side, and metallicity\cite{1,2},
well-defined Fermi surfaces\cite{14}, and BCS-like
superconductivity on the other\cite{4,5}. The under-doped region,
associated with a "pseudogap" state and a temperature scale
T*\cite{4,5}, has been identified through detailed changes in the
curvature of the normal state resistivity\cite{2}, changes in
specific heat jumps at Tc\cite{5}, or spectroscopic techniques, such as
optical conductivity\cite{4}, ARPES\cite{3, 14} (Angle-Resolved
Photoemission Spectroscopy) and Raman\cite{15}.  While superconductivity 
prevents the establishment of a putative hidden ordered phase 
underneath the superconducting dome via resistivity measurements, 
high field experiments that quench the superconductivity reveal an 
electronic transition underneath the superconducting dome\cite{1}.  
Muon spin resonance and Kerr rotation 
also suggest a critical point\cite{16, 17}.  However, the connection
of the critical point with high temperature superconductivity itself 
is missing.  With the exception of the 
recent work on the optical Kerr effect \cite{17},
changes in microscopic processes involving lattice and/or
magnetism have not been explicitly correlated with the evolution
of the electronic structure and scattering across optimal doping,
suggesting the need for a continuous tuning parameter to
investigate changes on a finer scale than doping can provide.

Through Raman spectroscopy and x-ray diffraction of a slightly
doped ($\delta$ $\sim$ 0.03), insulating parent compound of the
cuprates (B$i_{1.98}$Sr$_{2.06}$Y$_{0.68}$Cu$_{2}$O$_{8 + \delta}$) to
pressures of 35 GPa in a Diamond Anvil Cell (DAC), we establish a
critical pressure in a single sample and tie six anomalies in
electronic, spin, and lattice degrees of freedom to it. Pressure,
being a continuous, reversible, and laboratory controlled physical
variable, can tune a single sample across a phase transition.  As
a contact-less probe separable from its surroundings, Raman
spectroscopy circumvents the difficulties that prevent
conventional methods from investigating a critical point under
pressure. 

Therefore, we rely primarily on spectroscopic and diffraction data
to define the critical pressure.  We measure electronic
scattering, short-range magnetic correlation, and electron-phonon
coupling via the analysis of the electronic background and
collective mode frequencies/line-shape in Raman spectra\cite{15}.
By also performing x-ray diffraction experiments, we measure
compressibility of the lattice and phonons, identifying the c-axis
lattice density in these highly layered compounds as a
complimentary metric to pressure.

\begin{figure}[htb]
\includegraphics[width=0.5\textwidth]{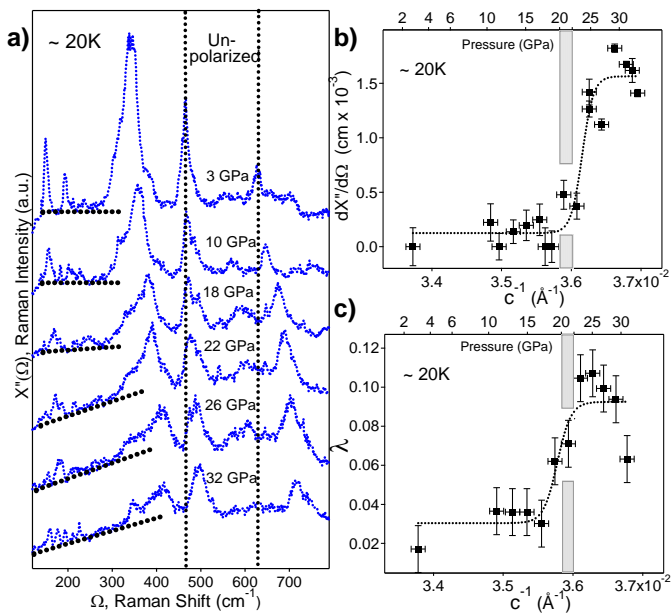}
\caption{\label{Fig1} a) Raman spectra including the B$_{1g}$ phonon, 
O-Bi phonon in the Bi-O block layer, and the apical oxygen phonon 
at $\sim$20K. b) Slope of the linear electronic background from 
200-300cm$^{-1}$ at $\sim$20K with an onset at 21 GPa.  c) The full
Raman vertex, using a normal state electronic background, was fit
to 20K data obtaining $\sim$ 0.1 at the higher pressures, and
exhibiting an onset at $\sim$ 21GPa. The procedure followed is
outlined in Ref. 24.}
\end{figure}

Single crystals of Bi$_{1.98}$Sr$_{2.06}$Y$_{0.68}$Cu$_{2}$O$_{8 +
\delta}$ were grown by the floating zone method, and have a doping
dependence described by Maeda\cite{18} and Terasaki\cite{19}.  Raman
spectroscopy was performed on $\sim$ 40x40 $\mu$m$^{2}$ crystals
in both back-scattering geometry with 514.5 nm light and
135$^{\circ}$ incidence geometry with 488 nm light of an Argon ion
laser. High-pressure structure determination was based on
angle-dispersive synchrotron powder x-ray diffractometry. The
diffraction experiments were carried out at the 16ID-B beamline of
the High Pressure Collaborative Access Team (HPCAT), Advanced
Photon Source of Argonne National Laboratory. The monochromatic
xray of wavelength $\sim$ 0.40165  \AA was focused to a beam size of
x $\sim$ 30 $\mu$m$^{2}$ with two bent mirrors. The beam was finally
reduced to a size of 10 $\mu$m with a pin-hole collimator and the
diffraction images were recorded with a MAR345 charge coupled
device and converted to 2$\Theta$ using Fit-2D \cite{20}. In both
Raman and x-ray diffraction experiments, samples were loaded into
a DAC containing a pressure chamber of a rhenium gasket filled
with a quasi-hydrostatic neon gas medium.  A ruby chip served as
an optical pressure sensor in the DAC\cite{21}.

Representative Raman spectra for the insulating compound are shown
in the low pressure spectra of Fig 1a.  The peaks at $\sim$
340cm$^{-1}$, $\sim$490cm$^{-1}$, and $\sim$620cm$^{-1}$ are,
respectively, the out-of-phase vibrations of the in-plane oxygen
(B$_{1g}$ "bond-buckling" mode), the oxygen vibration in the Bi-O
block layer, and the apical oxygen vibration \cite{22}. With
pressure, the oxygen vibration in the block layers far from the
Cu-O plane remains nearly constant, while the frequencies of the
B$_{1g}$ bond-buckling and the apical oxygen modes shift by over
$\sim$100 cm$^{-1}$. Pressure, then, predominantly modifies the
local crystal fields in the perovskite octahedral relevant to the
physics of the Cu-O plane.  In the pressure range shown, the
c-axis compresses by $\sim$ 10 $\%$ while the a,b-axes compress by
$\sim$ 5 $\%$ (see Fig 4).

\begin{figure}[htb]
\includegraphics[width=0.4\textwidth]{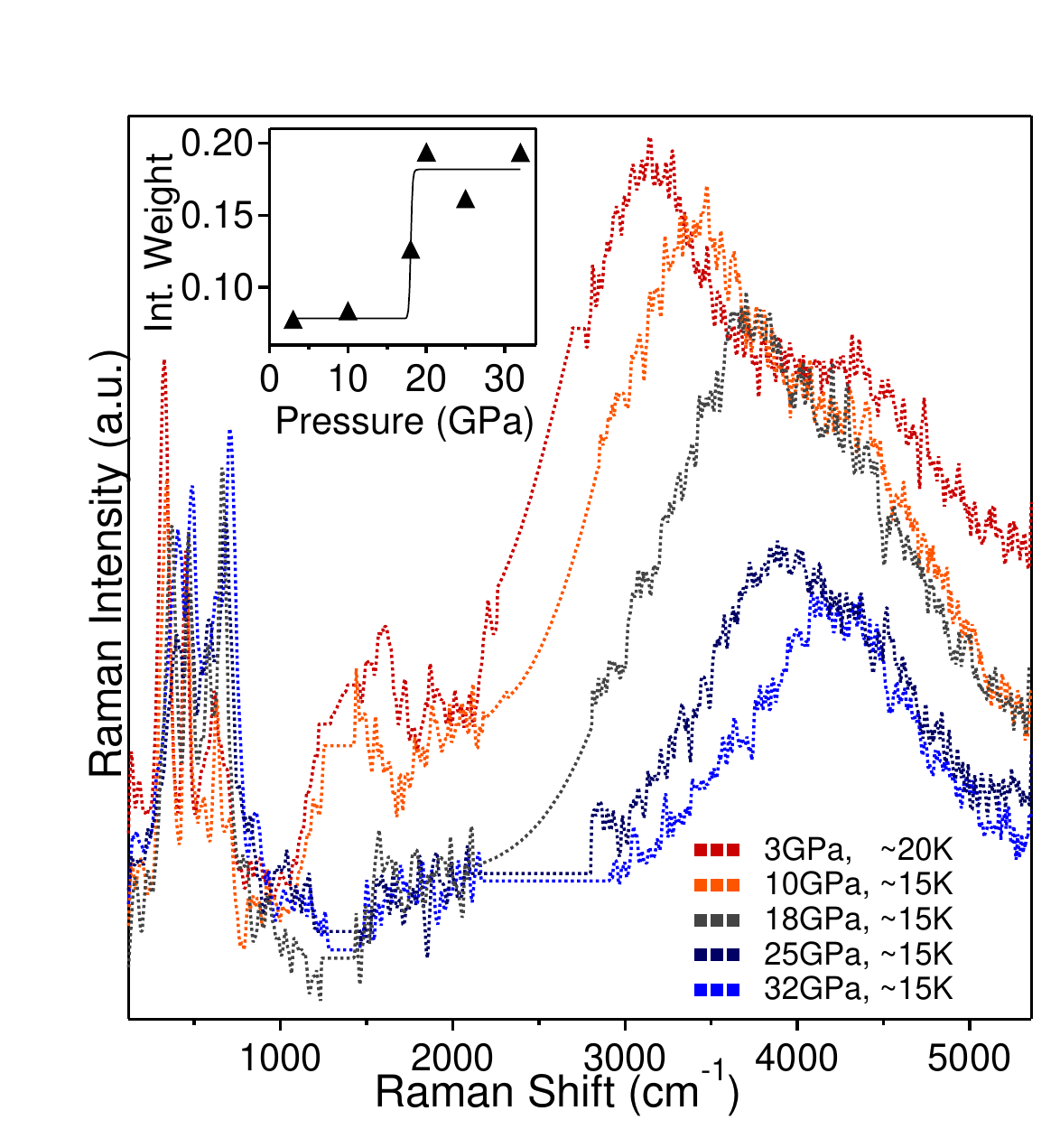}
\caption{\label{Fig2} Spectral weight decrease of the 
two-magnon peak ($\sim$3000-5000cm$^{-1}$) at $\sim$20K across the $\sim$21 GPa transition
coincides with an increase in spectral weight below
$\sim$1000cm$^{-1}$. Since the low frequency side of the two-magnon peak coincides with the
2$^{nd}$ order Raman peak of diamond, background spectra taken 
from the sample-free region were subtracted from the sample spectra. Inset graphs the
ratio of the integrated spectral weight below 1000cm$^{-1}$ to that above 1000cm$^{-1}$ for
each pressure point.}
\end{figure}

We establish a pressure driven electronic transition through the
onset of a linear electronic background and an abrupt increase of
electron-phonon coupling at $\sim$ 21 GPa in the low-T Raman
spectra of Fig 1.  The Raman cross-section may be related to the
conductivity in the following way:  $\chi"(\Omega) $   $\alpha$
  $\Omega\sigma'(\Omega)$, such that a change in conductivity
contributes a change of low frequency, linear background to the
Raman cross-section\cite{23,24}. Fig 1b shows the results of a
linear fit of the electronic background between $\sim$200cm$^{-1}$
and $\sim$ 300-350cm$^{-1}$---a spectral range below the oxygen
B$_{1g}$ phonon---with a sharp onset at $\sim$ 21GPa. Concomitant
to this change in the electronic background, the intensity of the
B$_{1g}$ phonon decreases significantly and its line-shape becomes
asymmetric. The full Fano line-shape of the B$_{1g}$ phonon was
fit using the theory of Ref. 25, giving a sharp increase of
$\lambda \sim$ 0.1 at $\sim$ 21 GPa\cite{25}.  The sharp change of
$\lambda$ is consistent with an increase of mobile charge carriers
in the presence of an unscreened electron-phonon
interaction\cite{26}. Upon further increase of metallicity, the
electron-phonon interaction becomes screened as in
conventional metals, consistent with the downward trend of
$\lambda$ beyond 25 GPa shown in Fig 1c. Thus both the electronic
background and Fano lineshape indicate an abrupt change in
conduction at $\sim$ 21 GPa.

\begin{figure*}[htb]
\includegraphics[width=1\textwidth] {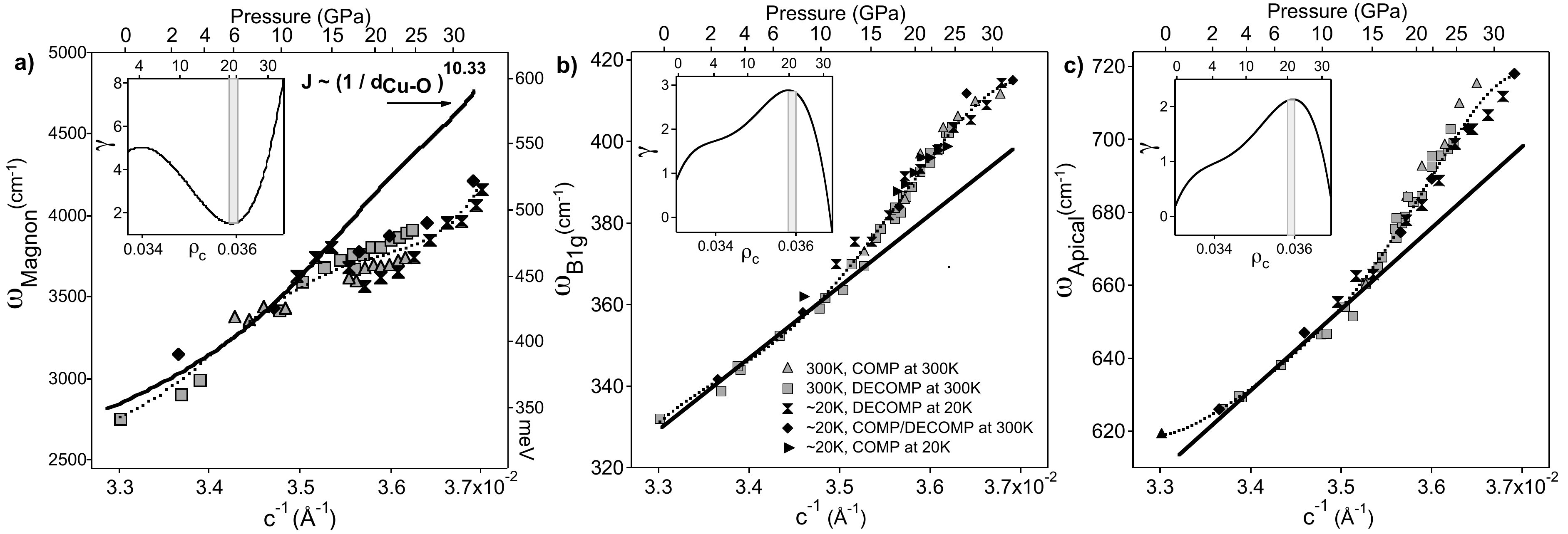}
\caption{\label{Fig3} Peak positions of the two-magnon peak (a), B$_{1g}$ phonon (b), and apical phonon (c),
at both low (20K) and high (300K) temperatures plotted against the
c-axis lattice density and pressure.  The data was taken using
several different pressure, temperature pathways indicated by the different
marker styles.  The two magnon peak data are compared with Heisenberg expectations where J $\sim$ (1/d$_{Cu-O})^{10.33}$) (solid line).  
The inset of (a) shows the derivative of a fit to the all data points (dotted curve) with a minimum at 21 GPa. 
For the phonons, such a derivative is expressed in the form of a Gruneisen parameter,
$\gamma$ $\sim$ dln($\omega$)/dln($\rho_{c}$), plotted in the
insets of (b) and (c) with a maximum at $\sim$21 GPa. Solid lines in (b) and (c) are guides to the eye, 
following the expected dependence at low pressures.}
\end{figure*}

We also find a weakening of the magnetic correlation by tracking the two-magnon peak, in analogy
to changes observed with doping\cite{27}.  
Data taken over a large spectral range reveal a substantial decrease in the
high frequency $\sim$ 3000 - 5000 cm$^{-1}$ two-magnon peak
spectral weight and concomitant spectral weight increase below
$\sim$ 1000cm$^{-1}$ across the $\sim$ 21 GPa transition (Fig 2).  
This spectral weight transfer suggests that the abrupt change
in conduction shown in Fig 1 occurs simultaneously with a weakening of the
magnetic correlation. Plotted in the inset of Fig 2 is the ratio of the integrated spectral
weight below 1,000 cm$^{-1}$ to that above 1,000 cm$^{-1}$ for each pressure point, showing a similar onset in the magnetic 
correlation at $\sim$ 21GPa.

Furthermore, the two-magnon peak softens significantly in the pressure range beyond the
electronic transition.  The two-magnon peak positions are plotted in 
Fig 3a versus pressure and the c-axis lattice density.  
In a canonical Mott-Hubbard insulator, such as K$_{2}$NiF$_{4}$, 
the experimentally determined\cite{28} scaling 
of the two-magnon peak is (1/d$_{Cu-O}$)$^{9.5}$, where d$_{Cu-O}$
is the copper-oxygen distance. We obtain a similar exponent of (1/d$_{Cu-O}$)$^{10.33}$
for the insulating, low pressure data. However, we find a
softening of $\sim$ 100meV in the pressure range beyond 21GPa from
the expected dependence.  Note that the x-ray diffraction data in
Fig 4 gives a relation between the a, b-axes and the c-axis that
allows one to plot the expected d$_{Cu-O}$ dependence versus
c-axis lattice density.

Coincident with the electronic transition at 21 GPa are also changes in 
lattice compressibility. Figs. 3b and 3c
show a maximum in the compressibility of two phonon modes
(B$_{1g}$ and apical oxygen).  The phonon
frequency of the B$_{1g}$ and apical oxygen mode scale
approximately linearly with c-axis lattice density up to $\sim$ 18
GPa, from 18-21GPa increase with a faster rate, and by 32 GPa show
saturation.   A derivative of a fit to all the points has a
maximum at $\sim$ 21GPa for both phonons (insets); this derivative
can be likened to a pressure-dependent Gruneisen parameter,
$\gamma$ ($\rho_{c}$) = dln($\omega$)/dln($\rho_{c}$). The same type of analysis was
done for the two-magnon mode (inset of Fig 3a) revealing, instead, a minimum
in the Gruneisen parameter at $\sim$ 21 GPa. 

\begin{figure}[htb]
\includegraphics[width=0.5\textwidth]{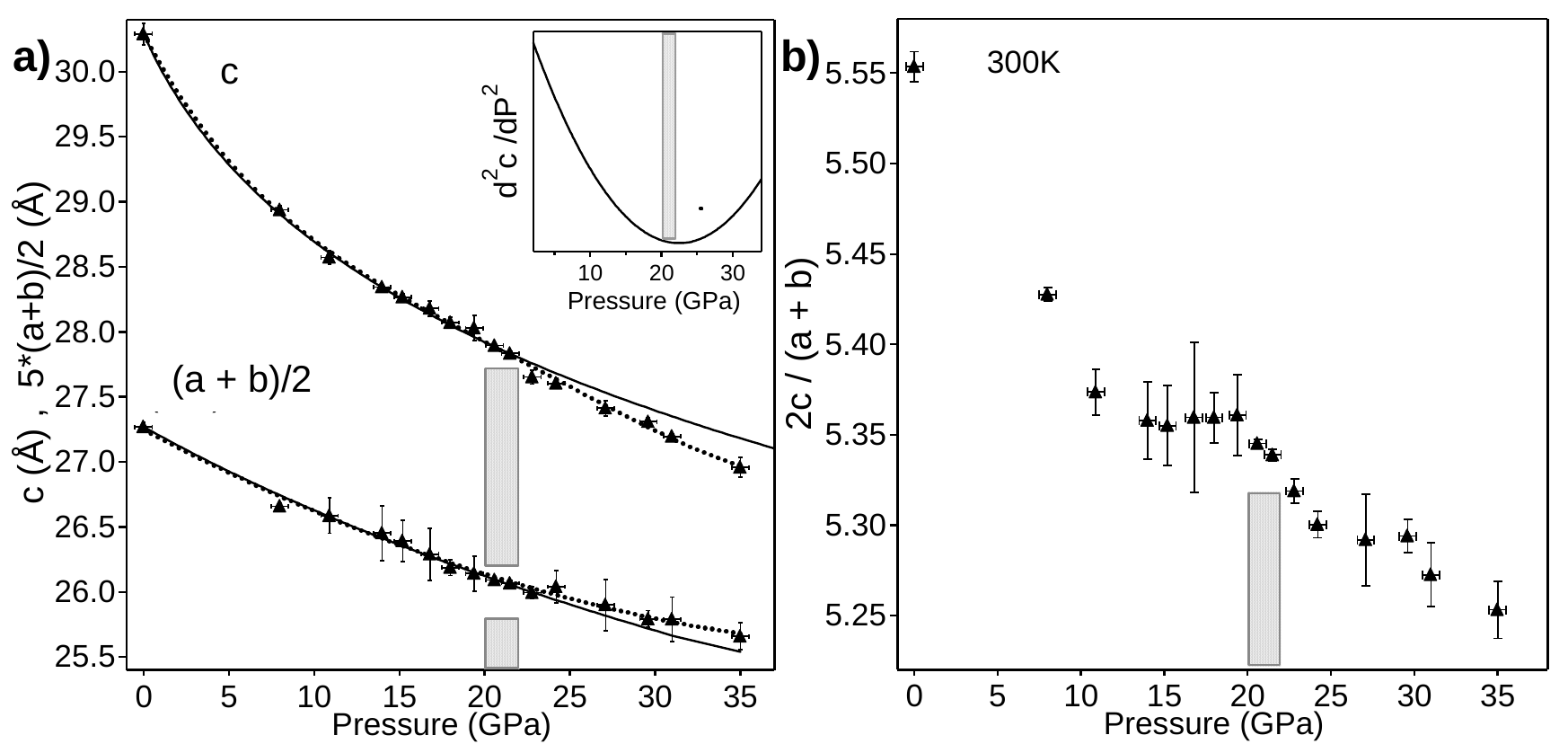}
\caption{\label{Fig4} a) The c-axis, and
5*(a+b)/2-axis lattice constants are plotted versus pressure. Solid lines are fits
to the Birch-Murnaghan equation of state between
0-20GPa. Dotted curves are polynomial fits to the entire data range between
0-35GPa.  A second derivative of this polynomial fit for the
c-axis is plotted in the inset with an inflection point
at 20-24GPa. b) Plot of 2c/(a+b) with an anomaly at 20-22GPa.}
\end{figure}

Fig 4a and 4b reveal a discontinuity
in the compressibility of the c-axis also at $\sim$ 21 GPa. A fit
to the Birch-Murnaghan equation of state deviates\cite{29} from
the c-axis x-ray data at 21 GPa.  This fit indicates a certain
bulk modulus, (1/V)$\delta$V/$\delta$P, below 21 GPa, and
different bulk modulus above.  Further, plots of c/a reveal a kink
at 21GPa; a second derivative of the polynomial fit to the c-axis
data set shows an inflection point at 20-24 GPa.

We now turn to how this critical pressure at 21 GPa resembles the
critical point at optimal doping and the associated
dichotomy in the phase diagram.  While due to the discrete doping
axis significantly fewer data points are plotted and with less of
a clear onset, Raman spectroscopy has also found an onset in the
electronic background occurring maximally at
$\sim$ 0.2 holes/Cu\cite{30}.  Further, the line-shape evolution under pressure
is not unlike that seen with doping.  As with higher pressures, an appreciable $\lambda$ 
is obtained for significantly doped compounds\cite{25, 31}. Finally, 
the two-magnon peak softens by a similar value ($\sim$ 150meV) 
from the insulating state to optimal doping\cite{27} as it does by 21GPa 
($\sim$100meV) here.  These comparisons suggest that $\sim$30 GPa 
is enough to traverse the high-Tc phase diagram, starting from 
a slightly doped parent insulator.    Screening may account for the
changes in  $\lambda$ and the exchange energy J across the phase diagram.  
With an abrupt increase in screening, the phonons should harden, 
accounting for the concavity of the phonon Gruneisen parameter, 
while the exchange energy should soften, accounting 
for the convexity of the magnon Gruneisen parameter.

One could also ascertain how the pressure driven transition
relates to optimal doping by investigating how pressure effects
more highly doped samples in the phase diagram. Past pressure
experiments on superconducting samples of
Bi$_{2}$Sr$_{2}$CaCu$_{2}$O$_{8+\delta}$ have shown increases in T$_{c}$ and
concomitant decreases in R$_{H}$, the Hall coefficient, on the
underdoped side of the dome, and decreases in T$_{c}$ with continued
decrease of R$_{H}$ on the overdoped side \cite{32,33}. The
predominant interpretation is that preferred compression along the
layered c-axis brings the chemically substituted block layers
closer to the Cu-O planes, inducing hole doping.  The dopant charge is 
not fully donated to the Cu-O plane in these compounds and therefore, hole count
depends on the distance of the Cu-O plane to the dopants.
Estimates of hole count based on LDA give $\sim$0.008 holes/Cu/GPa for both
underdoped and optimally doped samples\cite{34}, which would
suggest that a slightly doped parent insulator could reach optimal
doping (a hole count of $\sim$0.2 hole/Cu), by 20GPa.  However, this pressure induced
doping does not preclude pure electronic or lattice density from also driving the transition.  It does
suggest that the two axes-x, P are not independent tuning parameters.

In conclusion, we have established a critical
pressure at 21 GPa, where the electronic background and
electron-phonon coupling $\lambda$ show clear onsets of change,
spectral weight transfers from high to low frequencies, magnons
and phonons show maximum change in the Gruneisen parameter, and
the c-axis compressibility changes discontinuously.  This dramatic
result opens up the high-Tc field to a number of other condensed
matter experiments in the pressure domain.  It also suggests aspects of the 
data and ideas based on critical points inside the
superconducting dome\cite{1,2,3,4,5,6,7,8,9,10,11,12,13,14}. The
number of physical quantities showing anomalies across this
critical pressure indicates that a unifying low energy feature,
such as the Fermi surface, drives these changes. Further, the
independence of the results to a number of different P, T-
pathways and the discontinuity in the c-axis compressibility are
consistent with a continuous, 2nd order-like transition \cite{35}.
Therefore, we suggest an electronic Lifshitz transition \cite{36,
37} as a possible explanation.  A change in the topology of the
Fermi surface concomitant with a marked increase in mobile charge
carriers may naturally occur in the cuprates, for example when the van Hove
Singularity at ($\pi$,0) crosses the Fermi level. Such an
electronic transition would also be consistent with the doping
evolution seen in ARPES\cite{14} from pockets or arcs to a large
Fermi surface and with transport\cite{38} and optics\cite{39} data
where the carrier number crosses over from a small (x) to large
(1-x) Luttinger number near optimal doping.  However, using
pressure to continuously tune a single sample, we go beyond the
doping evolution by identifying a distinct transition at a
critical pressure and connecting lattice and magnetic anomalies to
it.

We would like to acknowledge helpful discussions with R.B.
Laughlin, S.A. Kivelson, S.C. Zhang, A. Mackenzie, E. Gregoryanz,
and O. Degatyerova. The work at Carnegie was supported by
DOE/NNSA and DOE/BES grant DE-FG02-02ER45955.  The Stanford work
was supported by DOE Office of Science, Division of Materials
Science, with contract DE-FG03-01ER45929-A001. TPD would like to 
acknowledge support from NSERC, CFI and the Alexander von Humboldt Foundation. 
CK acknowledges support from the ONR. Use of the HPCAT facility was supported by DOE-BES,
DOE-NNSA (CDAC), NSF, DOD -TACOM, and the W.M. Keck Foundation.
Use of the APS was supported by DOE-BES, under Contract No.
W-31-109-ENG-38.

\end{document}